# Fast Updating the STBC Decoder Matrices in the Uplink of a Massive MIMO System

**Seyed Hosein Mousavi**
*Department of Electrical & Computer Engineering*
*University of Tabriz*
*Tabriz, Iran*
*mir.hosein.mousavi@tabrizu.ac.ir*

**Jafar Pourrostam**
*Department of Electrical & Computer Engineering*
*University of Tabriz*
*Tabriz, Iran*
*j.pourrostam@tabrizu.ac.ir*

Abstract: *Reducing computational complexity of the modern wireless communication systems such as massive Multiple-Input Multiple-Output (MIMO) configurations is of utmost interest. In this paper, we propose new algorithm that can be used to accelerate matrix inversion in the decoding of space-time block codes (STBC) in the uplink of dynamic massive MIMO systems. A multi-user system in which the base station is equipped with a large number of antennas and each user has two antennas is considered. In addition, users can enter or exit the system dynamically. For a given space-time block coding/decoding scheme the computational complexity of the receiver will be significantly reduced when a user is added to or removed from the system by employing the proposed method. In the proposed scheme, the matrix inversion for zero-forcing (ZF) as well as minimum mean square error (MMSE) decoding is derived from the inverse of a partitioned matrix and the Woodbury matrix identity. Furthermore, the suggested technique can be utilized when the number of users is fixed but the channel estimate changes for a particular user. The mathematical equations for updating the inverse of the decoding matrices are derived and its complexity is compared to the direct way of computing the inverse. Evaluations confirm the effectiveness of the proposed approach.*

## I. Introduction

Massive MIMO, one of the underlying technologies for the new generations of wireless communication systems, has been explored in recent years [1]. In this scheme for cellular communications, the base station (BS) is equipped with a large number of antennas and simultaneously serves multiple users. In such configurations high capacity, energy efficiency as well as high reliability can be achieved via simple signal processing techniques [2]. Additionally, when the number of antennas at the BS is very large, the uplink communication channels will be asymptotically orthogonal. Therefore, intra-cell/inter-cell interference can be largely eliminated utilizing simple linear signal processing methods such as zero-forcing (ZF) or minimum mean square error (MMSE) [3]. In this fashion although multiple users transmit signals in the same frequency band and at the same time slots, virtual point-to-point SIMO links are established in which each user has single antenna and the BS has multiple antennas. Moreover, because the capacity of the multiple antenna systems is proportional to the minimum number of transmit and receive antennas [4], using one antenna in the transmitter will lower the overall throughput of the system.

One natural solution to improve the diversity gain of each user in uplink communications and increase the capacity of the system is using multiple antennas along with STBC at the transmitter (user) side [4]-[8]. It has been shown that by using a good space-time block code with full diversity and linear receiver, the intercellular interference problem can be solved to a high degree [4]. For a massive MIMO system with two antennas at the user terminal, in [4] the sufficient conditions to design a good STBC with linear receivers are studied. Furthermore its performance in terms of attainable throughput is investigated.

It is worth mentioning that many benefits of various massive MIMO configurations come at the price of high computational complexity. For example linear STBC decoding methods such as ZF and MMSE algorithms require inverting a matrix with large dimensions when the number of users increases. Therefore, computationally efficient methods must be developed to cope with this challenge and make the hardware implementation feasible. In terms of computing the inverse of a large matrix, there are two common approaches to tackle the complexity issue: (1) finding the exact inverse of a matrix by updating the available information about that matrix and using matrix inversion identities, and (2) calculating an estimate of the inverse for example using Neumann series [9]-[10]. In this paper we adapt the STBC scheme presented in [4] for a dynamic massive MIMO system. By dynamic we mean that users are entering or exiting the system. For the selected coding design, based on the matrix inversion lemmas such as the inverse of a partitioned matrix and the Woodbury formula [11], we propose and evaluate low-complexity methods to speed up STBC ZF and MMSE decoders. Update equations are derived for the cases that a user is added to or removed from the system as well as the case that the channel estimate of a user has changed. Algorithms are evaluated and compared in terms of computational complexity. The proposed algorithms have fewer computations which naturally lead to reduction in the running time of a Software-Defined-Radio (SDR) program or complexity of implemented hardware for the receiver in the system. Not only can these algorithms be used in a slow fading environment by switching active users, but also could be used in fast fading channels with frequent changes to the user channel estimates.

The paper is organized as follows. The system model and the STBC scheme and corresponding decoder methods are described in Section II. In Section III, the fast matrix inversion equations are derived for each of the above mentioned scenarios. The computational complexity of the suggested approaches is investigated in Section IV and conclusions are drawn in Section V.

## II. SYSTEM MODEL

Consider the uplink of a cellular multi-user massive MIMO system in which the BS is equipped with $N$ antennas and serves $M$ user ($M < N$) in such a way that each independent user has two antennas, as illustrated in Figure 1. The channel is supposed to follow Rayleigh small fading and large scale path loss and shadowing model.

The channel gain between the $p$-th antenna of the $m$-th user and the $n$-th antenna of the BS is formulated as $\beta_{nmp} h_{nmp}$ ($1 \leq n \leq N, 1 \leq m \leq M, p = 1,2$), where $\beta_{nmp}$ is related to the large scale path loss and shadowing and $h_{nmp}$ shows the small scale fading. It is assumed that $\beta_{nmp} = \beta_m$ for $n = 1, \dots, N$ and $p = 1,2$. In addition, to normalize the average power, we assume that $\beta_1 = 1$ and $\beta_1 \geq \beta_2 \geq \dots \geq \beta_M$. Based on the Rayleigh fading model, $h_{nmp}$ is assumed to be an independent and identically distributed (i.i.d) zeromean, circularly symmetric complex Gaussian random variable with unit variance. Furthermore, the fading coefficients and the large scale channel gains from the $m$-th user to the BS are expressed as $\mathbf{H}_m = [h_{nmp}]_{N \times 2}$ and $\mathbf{L}_m = \beta_m \mathbf{I}_2$ respectively.

Suppose STBC is adopted by each subscriber in the cell, and we express the code of the $m$-th user as $\widetilde{\mathbf{X}}_m$ matrix with the size of $2 \times S$. With these assumptions the received signals in the base station over $S$ time slots, $\mathbf{Y}_{N \times S}$, is written as follows:

$$\mathbf{Y} = \sum_{m=1}^{M} \sqrt{\frac{\rho}{2}} \mathbf{H}_m \mathbf{L}_m \widetilde{\mathbf{X}}_m + \mathbf{W} = \sqrt{\frac{\rho}{2}} \mathbf{H} \mathbf{L} \widetilde{\mathbf{X}} + \mathbf{W} \quad (1)$$

where $\mathbf{H} = [\mathbf{H}_1 \quad \mathbf{H}_2 \quad \dots \quad \mathbf{H}_M]$, $\mathbf{L} = diag(\mathbf{L}_1, \mathbf{L}_2, \dots, \mathbf{L}_M)$ is a block diagonal matrix, and $\widetilde{\mathbf{X}} = [\widetilde{\mathbf{X}}_1^T, \widetilde{\mathbf{X}}_2^T, \dots, \widetilde{\mathbf{X}}_M^T]^T$ with energy restriction $E\{tr(\widetilde{\mathbf{X}} \widetilde{\mathbf{X}}^H)\} = 2S$, superscripts $T$ and $H$ represent the matrix Transpose and Hermitian operators respectively. Also, $\rho$ demonstrates the received SNR and $\sqrt{1/2}$ is used to nomalize the transmit signal energy to be "1" per time slot. $\mathbf{W}_{N \times S}$ represents the noise whose entries are i.i.d. taken from the zero-mean, circularly-symmetric complex Gaussian random variables with unit variance.

Next we explain the STBC coding and decoding algorithms.

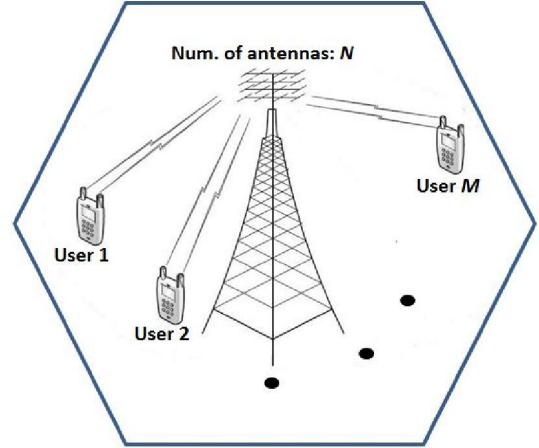

Figure 1. A Cellular Multiuser Massive MIMO System

### A. Coding Matrix for each User

The transmitted signal matrix $\widetilde{\mathbf{X}}$ is an STBC which is sent from two transmit antennas over $S$ time slots. In this paper, we choose $S = 2$ and the corresponding STBCs for the $m$-th user $\widetilde{\mathbf{X}}_m$ is designed as follows, i.e.,

$$\widetilde{\mathbf{X}}_m = \begin{bmatrix} a_m(x_{m1} + b_m x_{m2}) & \gamma_m a_m(x_{m3} + b_m x_{m4}) \\ c_m(x_{m3} + d_m x_{m4}) & c_m(x_{m1} + d_m x_{m2}) \end{bmatrix} \quad (2)$$

where $\mathbf{x}_m = [x_{m1}, x_{m2}, x_{m3}, x_{m4}]^T$ transmitted symbols vector of $m$-th user and $a_m, b_m, c_m, d_m, \gamma_m$ are constants that are determined to satisfy the orthogonality of the code.

Denoting $\mathbf{H}_m = [\mathbf{h}_{m1} \quad \mathbf{h}_{m2}]$ and substituting (2) into (1) and taking vector form of $\mathbf{Y}$ we have:

$$vec(\mathbf{Y}) = \sum_{m=1}^{M} \sqrt{\frac{\rho}{2}} \beta_m \widetilde{\mathbf{H}}_m \mathbf{x}_m + vec(\mathbf{W}) \quad (3)$$

where

$$\widetilde{\mathbf{H}}_m = \begin{bmatrix} a_m \mathbf{h}_{m1} & a_m b_m \mathbf{h}_{m1} & c_m \mathbf{h}_{m2} & c_m d_m \mathbf{h}_{m2} \\ c_m \mathbf{h}_{m2} & c_m d_m \mathbf{h}_{m2} & \gamma_m a_m \mathbf{h}_{m2} & \gamma_m a_m b_m \mathbf{h}_{m2} \end{bmatrix} \quad (4)$$

For linear decoders such as ZF and MMSE filters, it is desired that the columns of $\widetilde{\mathbf{H}}_m$, when $N$ is large enough, are asymptotically orthogonal. Applying the orthogonality and the energy ($E\{tr(\widetilde{\mathbf{X}} \widetilde{\mathbf{X}}^H)\} = 4$) criteria, it is shown that the coding constants are obtained as follows [4]:

$$a_m = (1 + \mathbf{j}(1 - b_m))/\sqrt{5} \quad , \quad b_m = (1 + \sqrt{5})/2$$
$$c_m = (1 + \mathbf{j}(1 - d_m))/\sqrt{5} \quad , \quad d_m = (1 - \sqrt{5})/2$$
$$\gamma_m = \mathbf{j} \quad (5)$$

### B. Linear Decoding for each User

We define $\widetilde{\mathbf{G}}$ as a matrix with dimensions of $2N \times 4M$ and $\mathbf{x}$ as a $4M$-dimentional vector:

$$\widetilde{\mathbf{G}} = [\beta_1 \widetilde{\mathbf{H}}_1 \quad \beta_2 \widetilde{\mathbf{H}}_2 \quad \ldots \quad \beta_M \widetilde{\mathbf{H}}_M]$$
$$\mathbf{x} = [\mathbf{x}_1^T, \quad \mathbf{x}_2^T, \quad \ldots, \mathbf{x}_M^T]^T \quad (6)$$

Hence, we can rewrite equation (3) as:

$$vec(\mathbf{Y}) = \sqrt{\frac{\rho}{2}} \widetilde{\mathbf{G}} \mathbf{x} + vec(\mathbf{W}) \quad (7)$$

Let $\mathbf{Q}_{ZF}$ and $\mathbf{Q}_{MMSE}$ be tbohe ZF decoder and the MMSE decoder matrices, respectively, we will have:

$$\mathbf{Q}_{ZF} = (\widetilde{\mathbf{G}}^H \widetilde{\mathbf{G}})^{-1} \widetilde{\mathbf{G}}^H \quad (8)$$

$$\mathbf{Q}_{MMSE} = (\frac{2\mathbf{I}_{4M}}{\rho} + \widetilde{\mathbf{G}}^H \widetilde{\mathbf{G}})^{-1} \widetilde{\mathbf{G}}^H \quad (9)$$

Multiplying (7) by these matrices from the left, we have:

$$\mathbf{Q} vec(\mathbf{Y}) = \sqrt{\frac{\rho}{2}} \mathbf{Q} \widetilde{\mathbf{G}} \mathbf{x} + \mathbf{Q} vec(\mathbf{W}) \quad (10)$$

where $\mathbf{Q} = \mathbf{Q}_{ZF}$ or $\mathbf{Q} = \mathbf{Q}_{MMSE}$. The $l$-th transmitted symbol of the $m$-th user $\hat{x}_{ml}$ is estimated as:

$$\hat{x}_{ml} = \underset{x}{argmin} \left| [\mathbf{Q} vec(\mathbf{Y})]_{4(m-1)+l} - \sqrt{\frac{\rho}{2}} \tilde{q}_{4(m-1)+l, 4(m-1)+l}\, x \right|, l = 1,2,3,4 \quad (11)$$

where the minimum is over the signal constellation of the $m$-th user, $[\mathbf{Q} vec(\mathbf{Y})]_p$ is the $p$-th element of the vector $\mathbf{Q} vec(\mathbf{Y})$, and $\tilde{q}_{u,v} = [\mathbf{Q} \widetilde{\mathbf{G}}]_{u,v}$.

### III. INVERSE MATRIX UPDATING

From (8) and (9), it can be seen that the decoding of the received signal involves computing the inverse of the following matrices:

$$\mathbf{Z} = \begin{cases} \widetilde{\mathbf{G}}^H \widetilde{\mathbf{G}} & for \text{ ZF} \\ \frac{2\mathbf{I}_{4M}}{\rho} + \widetilde{\mathbf{G}}^H \widetilde{\mathbf{G}} & for \text{ MMSE} \end{cases} \quad (12)$$

where $\widetilde{\mathbf{G}}$ is a $2N \times 4M$ dimensional matrix which contains the coding constants as well as the channel coefficients, however, from now on, for brevity it will be called the channel matrix. In this section we derive the fast matrix inverse update equations to reduce the computational complexity of the decoder for the cases that a user is added to or removed from the system and when the Channel State Information (CSI) of a particular user varies as well. The proposed solutions are based on the inverse of a partitioned matrix and the Woodbury matrix identity. Suppose matrix $\mathbf{Z}$ is partitioned as:

$$\mathbf{Z} = \begin{bmatrix} \mathbf{A} & \mathbf{B} \\ \mathbf{C} & \mathbf{D} \end{bmatrix} \quad (13)$$

where $\mathbf{A}$ and $\mathbf{D}$ are square matrices, the inverse of $\mathbf{Z}$ is given as:

$$\mathbf{Z}^{-1} = \begin{bmatrix} \mathbf{A} & \mathbf{B} \\ \mathbf{C} & \mathbf{D} \end{bmatrix}^{-1} = \begin{bmatrix} \mathbf{F}_{11} & \mathbf{F}_{12} \\ \mathbf{F}_{21} & \mathbf{F}_{22} \end{bmatrix} \quad (14)$$

where

$$\mathbf{F}_{11} = (\mathbf{A} - \mathbf{B}\mathbf{D}^{-1}\mathbf{C})^{-1}$$
$$\mathbf{F}_{12} = -\mathbf{F}_{11}\mathbf{B}\mathbf{D}^{-1}$$
$$\mathbf{F}_{21} = -\mathbf{D}^{-1}\mathbf{C}\mathbf{F}_{11}$$
$$\mathbf{F}_{22} = (\mathbf{D} - \mathbf{C}\mathbf{A}^{-1}\mathbf{B})^{-1} \quad (15)$$

In addition using the Woodbury formula, we have:

$$(\mathbf{A} - \mathbf{B}\mathbf{D}^{-1}\mathbf{C})^{-1} = \mathbf{A}^{-1} + \mathbf{A}^{-1}\mathbf{B}(\mathbf{D} - \mathbf{C}\mathbf{A}^{-1}\mathbf{B})^{-1}\mathbf{C}\mathbf{A}^{-1} \quad (16)$$

Hence, equations given in (15) can be equivalently written as:

$$\mathbf{F}_{22} = (\mathbf{D} - \mathbf{C}\mathbf{A}^{-1}\mathbf{B})^{-1}$$
$$\mathbf{F}_{11} = \mathbf{A}^{-1} + \mathbf{A}^{-1}\mathbf{B}\mathbf{F}_{22}\mathbf{C}\mathbf{A}^{-1}$$
$$\mathbf{F}_{12} = -\mathbf{A}^{-1}\mathbf{B}\mathbf{F}_{22}$$
$$\mathbf{F}_{21} = -\mathbf{F}_{22}\mathbf{C}\mathbf{A}^{-1} \quad (17)$$

Next the algorithms for updating both ZF and MMSE decoder matrices are described for different scenarios.

### A. Adding a User

First, we will examine the case where a user is added to the system. Suppose that initial channel matrix is $[\widetilde{\mathbf{G}}]_{2N \times 4M}$, and let the channel matrix of the user which enters the system be $[\mathbf{G}_a]_{2N \times 4}$. Now, define the new inflated matrix as $\mathbf{G}_e = [\widetilde{\mathbf{G}} \quad \mathbf{G}_a]$. Thus, the ZF decoding matrix defined in (12) is given as:

$$\mathbf{Z}_{e(ZF)} = \mathbf{G}_e^H \mathbf{G}_e = \begin{bmatrix} \widetilde{\mathbf{G}}^H \\ \mathbf{G}_a^H \end{bmatrix} [\widetilde{\mathbf{G}} \quad \mathbf{G}_a]$$
$$= \begin{bmatrix} \widetilde{\mathbf{G}}^H \widetilde{\mathbf{G}} & \widetilde{\mathbf{G}}^H \mathbf{G}_a \\ \mathbf{G}_a^H \widetilde{\mathbf{G}} & \mathbf{G}_a^H \mathbf{G}_a \end{bmatrix} \quad (18)$$

Thus, the resulting matrix has a dimension of $4(M+1) \times 4(M+1)$. Using (17), the inverse of the decoding matrix is calculated as follows:

$$\mathbf{Z}_{e(ZF)}^{-1} = \begin{bmatrix} \widetilde{\mathbf{G}}^H\widetilde{\mathbf{G}} & \widetilde{\mathbf{G}}^H\mathbf{G}_a \\ \mathbf{G}_a^H\widetilde{\mathbf{G}} & \mathbf{G}_a^H\mathbf{G}_a \end{bmatrix}^{-1} = \begin{bmatrix} \mathbf{F}_{11} & \mathbf{F}_{12} \\ \mathbf{F}_{21} & \mathbf{F}_{22} \end{bmatrix} \quad (19)$$

where

$$\mathbf{F}_{22} = (\mathbf{G}_a^H\mathbf{G}_a - \mathbf{B}^H\mathbf{Z}_{o(ZF)}^{-1}\mathbf{B})^{-1}$$
$$\mathbf{F}_{11} = \mathbf{Z}_{o(ZF)}^{-1} + \mathbf{Z}_{o(ZF)}^{-1}\mathbf{B}\mathbf{F}_{22}\mathbf{B}^H\mathbf{Z}_{o(ZF)}^{-1}$$
$$\mathbf{F}_{12} = -\mathbf{Z}_{o(ZF)}^{-1}\mathbf{B}\mathbf{F}_{22}$$
$$\mathbf{F}_{21} = -\mathbf{F}_{22}\mathbf{B}^H\mathbf{Z}_{o(ZF)}^{-1} \quad (20)$$

where $\mathbf{Z}_{o(ZF)}^{-1} = [\widetilde{\mathbf{G}}^H\widetilde{\mathbf{G}}]^{-1}$ is the inverse matrix before updating and $\mathbf{B} = [\widetilde{\mathbf{G}}^H\mathbf{G}_a]_{4M\times 4}$. As it is observed, in this algorithm we only need to calculate the inverse of $\mathbf{F}_{22}$ which is a $4\times 4$ dimensional matrix and the rest of the computations are matrix multiplication. However, direct calculation the decoding matrix needs the inversion of a matrix with dimensions of $4(M+1) \times 4(M+1)$. Similarly, for the MMSE decoder we have:

$$\mathbf{Z}_{e(MMSE)} = \begin{bmatrix} \widetilde{\mathbf{G}}^H\widetilde{\mathbf{G}} + (2/\rho)\mathbf{I}_{4M} & \widetilde{\mathbf{G}}^H\mathbf{G}_a \\ \mathbf{G}_a^H\widetilde{\mathbf{G}} & \mathbf{G}_a^H\mathbf{G}_a + (2/\rho)\mathbf{I}_4 \end{bmatrix} \quad (21)$$

and

$$\mathbf{Z}_{e(MMSE)}^{-1} = \begin{bmatrix} \mathbf{F}_{11} & \mathbf{F}_{12} \\ \mathbf{F}_{21} & \mathbf{F}_{22} \end{bmatrix}$$
$$\mathbf{F}_{22} = (\mathbf{D} - \mathbf{B}^H\mathbf{Z}_{o(MMSE)}^{-1}\mathbf{B})^{-1}$$
$$\mathbf{F}_{11} = \mathbf{Z}_{o(MMSE)}^{-1} + \mathbf{Z}_{o(MMSE)}^{-1}\mathbf{B}\mathbf{F}_{22}\mathbf{B}^H\mathbf{Z}_{o(MMSE)}^{-1}$$
$$\mathbf{F}_{12} = -\mathbf{Z}_{o(MMSE)}^{-1}\mathbf{B}\mathbf{F}_{22}$$
$$\mathbf{F}_{21} = -\mathbf{F}_{22}\mathbf{B}^H\mathbf{Z}_{o(MMSE)}^{-1} \quad (22)$$

where $\mathbf{Z}_{o(MMSE)}^{-1} = (\widetilde{\mathbf{G}}^H\widetilde{\mathbf{G}} + (2/\rho)\mathbf{I}_{4M})^{-1}$ is the inverse matrix before updating, $\mathbf{B} = [\widetilde{\mathbf{G}}^H\mathbf{G}_a]_{4M\times 4}$, and $\mathbf{D} = \mathbf{G}_a^H\mathbf{G}_a + (2/\rho)\mathbf{I}_4$.

*B. Removing a user*

Now we consider a scenario in which a user is removed from the system and we break the current channel matrix as $\widetilde{\mathbf{G}} = [\mathbf{G}_f \quad \mathbf{G}_r]$ where $\mathbf{G}_r$ is the channel matrix of the user to be removed and $\mathbf{G}_f$ is the channel matrix after the removal of the user. In this case the update of the ZF decoding matrix involves calculating $\mathbf{Z}_{f(ZF)}^{-1} = (\mathbf{G}_f^H\mathbf{G}_f)^{-1}$. Using the inverse of a partitioned matrix, before the user is removed we have:

$$\mathbf{Z}_{o(ZF)}^{-1} = [\widetilde{\mathbf{G}}^H\widetilde{\mathbf{G}}]^{-1} = \begin{bmatrix} \mathbf{G}_f^H\mathbf{G}_f & \mathbf{G}_f^H\mathbf{G}_r \\ \mathbf{G}_r^H\mathbf{G}_f & \mathbf{G}_r^H\mathbf{G}_r \end{bmatrix}^{-1}$$
$$= \begin{bmatrix} \mathbf{F}_{11} & \mathbf{F}_{12} \\ \mathbf{F}_{21} & \mathbf{F}_{22} \end{bmatrix} \quad (23)$$
$$= \begin{bmatrix} (\mathbf{G}_f^H\mathbf{G}_f)^{-1} + (\mathbf{G}_f^H\mathbf{G}_f)^{-1}\mathbf{B}\mathbf{F}_{22}\mathbf{B}^H(\mathbf{G}_f^H\mathbf{G}_f)^{-1} & -(\mathbf{G}_f^H\mathbf{G}_f)^{-1}\mathbf{B}\mathbf{F}_{22} \\ -\mathbf{F}_{22}\mathbf{B}^H(\mathbf{G}_f^H\mathbf{G}_f)^{-1} & \mathbf{F}_{22} \end{bmatrix}$$

where $\mathbf{B} = \mathbf{G}_f^H\mathbf{G}_r$, hence we can write:

$$\mathbf{F}_{11} = \mathbf{Z}_{f(ZF)}^{-1} + \mathbf{Z}_{f(ZF)}^{-1}\mathbf{B}\mathbf{F}_{22}\mathbf{B}^H\mathbf{Z}_{f(ZF)}^{-1}$$
$$= \mathbf{Z}_{f(ZF)}^{-1} + \mathbf{F}_{12}\mathbf{F}_{22}^{-1}\mathbf{F}_{21} \quad (24)$$

which means to update the inverse of the ZF decoding matrix, we partition the current inverse and find the updated inverse as:

$$\mathbf{Z}_{f(ZF)}^{-1} = \mathbf{F}_{11} - \mathbf{F}_{12}\mathbf{F}_{22}^{-1}\mathbf{F}_{21} \quad (25)$$

Also, for the MMSE decoder we need to compute $\mathbf{Z}_{f(MMSE)}^{-1} = (\mathbf{G}_f^H\mathbf{G}_f + (2/\rho)\mathbf{I}_{4(M-1)})^{-1}$. Before the user is removed we have:

$$\mathbf{Z}_{o(MMSE)}^{-1} = \begin{bmatrix} \mathbf{F}_{11} & \mathbf{F}_{12} \\ \mathbf{F}_{21} & \mathbf{F}_{22} \end{bmatrix}$$
$$= \begin{bmatrix} \mathbf{G}_f^H\mathbf{G}_f + (2/\rho)\mathbf{I}_{4(M-1)} & \mathbf{G}_f^H\mathbf{G}_r \\ \mathbf{G}_r^H\mathbf{G}_f & \mathbf{G}_r^H\mathbf{G}_r + (2/\rho)\mathbf{I}_4 \end{bmatrix}^{-1} \quad (26)$$

Therefore similar to what we derived for the ZF decoder, we have:

$$\mathbf{Z}_{f(MMSE)}^{-1} = \mathbf{F}_{11} - \mathbf{F}_{12}\mathbf{F}_{22}^{-1}\mathbf{F}_{21} \quad (27)$$

where $\mathbf{F}_{11}, \mathbf{F}_{12}, \mathbf{F}_{21}$, and $\mathbf{F}_{22}$ are obtained from partitioning the current inverse matrix.

*C. Updating a user*

When new channel estimate is obtained (CSI is updated) for a particular user, the number of rows and columns of the channel matrix remains the same. For this case we propose an algorithm consisting two major steps. In this algorithm first, we delete the rows and columns of the updated user utilizing the proposed algorithm for removing a user. Then, using the proposed algorithm for adding a user we apply the new channel coefficients and update the inverse matrix. In other words, in ZF decoding, we first use equations (24) and (25) to remove the rows and columns of the particular user. Then, we use (19) and (20) for the final update of the inverse matrix. Clearly, for the MMSE decoder equations (26) and (27) are used first, and in the second step (22) is applied to find the inverse of the updated channel matrix.

## IV. COMPLEXITY ANALYSIS

In this section we evaluate and compare the computational complexity of the ZF STBC decoder in the uplink of a massive MIMO system with $N = 100$ antennas, and $M = 10, 16, 24, 30$ users. Note that similar results are applicable for the case of MMSE decoder. The computational complexity is studied in terms of the number of arithmetic operations. Assuming that the matrix whose inverse needs to be updated is $4M\times 4M$ dimensional, and K is the number of rows and columns that are added to or deleted from the matrix, the number of computations needed for decoder matrix inversion is summarized in Table I. In this table inflated channel matrix refers to the case that a user is added to the system $M_{new} = M + 1$ and deflated matrix represents the case that a user is removed from the system $M_{new} = M - 1$. It is clear that for the signal model and the STBC scheme used in this paper K = 4. The 4th row of Table I. corresponds to the case in which a new channel estimate is obtained for a particular user.

For the case that a user enters the system the complexity of the decoder is compared for different number of users and different methods of inverse matrix calculation in Table II. Moreover in Table III., the complexity reduction is compared when a user exits the system. Table IV compares the computational complexity of the two-stage update algorithm with the exact algorithm. As it can be seen, applying the proposed update techniques for the inverse matrix calculation results in considerable reduction in the computational complexity of the system. In addition as the number of current users in the system increases the complexity reduction gets bigger.

TABLE I. COMPUTATIONAL COMPLEXITY OF DIFFERENT ALGORITHMS

| | Computational complexity |
|---|---|
| Direct inverse | $(4(4M_{new})^3 + 10(4M_{new})^2 - 7(4M_{new}))/6$ |
| Update inflated | $K^3 + K^2(12M + 1) + (4K + 1)(4M)^2$ |
| Update deflated | $K^3 + K^2(4M) + (K + 1)(4M)^2$ |
| Updating new CSI | $2K^3 + K^2(16M - 3) + (K + 1)(4(M - 1))^2 + (4K + 1)(4M)^2$ |

TABLE II. ESTIMATED NUMBER OF OPERATIONS REQUIRED FOR CALCULATION OF INVERSE MATRIX: A USER IS ADDED TO THE SYSTEM

| Number of users ($M$) | 10 | 16 | 24 | 30 |
|---|---|---|---|---|
| Direct inverse | 59965 | 217250 | 683220 | 1296600 |
| Proposed inverse | 29200 | 72784 | 161360 | 250640 |
| Complexity reduction | 51% | 66% | 77% | 81% |

TABLE III. ESTIMATED NUMBER OF OPERATIONS REQUIRED FOR CALCULATION OF INVERSE MATRIX: A USER IS REMOVED FROM THE SYSTEM

| Number of users ($M$) | 10 | 16 | 24 | 30 |
|---|---|---|---|---|
| Direct inverse | 33200 | 149930 | 533120 | 1062900 |
| Proposed inverse | 8704 | 21568 | 47680 | 73984 |
| Complexity reduction | 74% | 86% | 91% | 93% |

TABLE IV. ESTIMATED NUMBER OF OPERATIONS REQUIRED FOR CALCULATION OF INVERSE MATRIX: NEW CHANNEL ESTIMATION FOR A PARTICULAR USER

| Number of users ($M$) | 10 | 16 | 24 | 30 |
|---|---|---|---|---|
| Direct inverse | 45287 | 181510 | 605072 | 1175860 |
| Proposed inverse | 36256 | 91744 | 205152 | 319776 |
| Complexity reduction | 20% | 49% | 66% | 73% |

## V. CONCLUSIONS

In this paper, the problem of fast matrix inversion is studied to reduce the computational complexity of linear STBC decoders in the uplink of a massive MIMO wireless communication system. Utilizing matrix inverse identities, efficient algorithms are proposed to update the inverse matrix for ZF and MMSE decoders when users are entering or exiting the system. Complexity analysis and evaluations demonstrate the effectiveness of the suggested methods. It is clear that the performance of the decoder remains the same when using the update algorithms. It is worth mentioning that similar approach will also be applicable when more users are added to or removed from the system. However, as the number of users to be added to or eliminated from the system increases, the gain of reduction in computational complexity decreases.